# Pinpointing the Dominant Component of Contact Resistance to Atomically Thin Semiconductors

Emanuel Ber[1], Ryan W. Grady[2], Eric Pop[2,3], and Eilam Yalon[1,*]

[1]*Viterbi Faculty of Electrical & Computer Engineering, Technion-Israel Institute of Technology, Haifa-32000, Israel*
[2]*Department of Electrical Engineering, Stanford University, Stanford, CA 94305, USA*
[3]*Department of Materials Science & Engineering, Stanford University, Stanford, CA 94305, USA*
[*]*E-mail:* eilamy@technion.ac.il

**Abstract**

Achieving good electrical contacts is one of the major challenges in realizing devices based on atomically thin two-dimensional (2D) semiconductors. Several studies have examined this hurdle, but a universal understanding of the contact resistance and an underlying approach to its reduction are currently lacking. In this work we expose the shortcomings of the classical contact resistance model in describing contacts to 2D materials, and offer a correction based on the addition of a lateral pseudo-junction resistance component ($R_{\text{jun}}$). We use a combination of unique contact resistance measurements to experimentally characterize $R_{\text{jun}}$ for Ni contacts to monolayer $MoS_2$. We find that $R_{\text{jun}}$ is the dominating component of the contact resistance in undoped 2D devices and show that it is responsible for most of the back-gate bias and temperature dependence. Our corrected model and experimental results help understand the underlying physics of state-of-the-art contact engineering approaches in the context of minimizing $R_{\text{jun}}$.

For five decades silicon-based complementary metal-oxide-semiconductor (CMOS) technologies have relied on reducing the size of transistors to increase their density on a chip, enabling a steady increase in computation capability.[1,2] As the lateral dimensions of field effect transistors (FETs) decrease, other dimensions must follow suit to retain electrostatic gate control.[3,4] However, reducing the thickness of transistor channels below ~4 nm raises difficulties for conventional bulk semiconductors.[4] Two-dimensional (2D) semiconductors are therefore considered prime candidates for ultrathin channels due to their few-atom thickness, alongside other advantages (e.g., heterogenous integration, substrate independence, and more).[4–6] Among the main difficulties in realizing FETs based on 2D materials is the high contact resistance ($R_c$) which limits the FET current.[7] Recent works have put great effort to reduce $R_c$ using approaches such as varying the contact material and access region doping.[8–15] However, how each approach affects the various components of the measured $R_c$ remains poorly understood. A universal method to characterize the contact resistance bottleneck is key for matching the contact resistance reduction method and future contact engineering.

Classically, an electrical contact between a metal and semiconductor is described by its $R_c$, which depends on the specific contact resistivity ($\rho_c$) and the sheet resistance under the contact ($R_{sk}$), as seen in Figure 1a.[16] Note that $R_{sk}$ is generally different from the sheet resistance in the channel ($R_{sh}$), stemming from the various contact formation processes (e.g., silicidation, doping, contact deposition) that can affect the region under the contact.[17] These processes are expected to have a more severe impact on atomically thin 2D FETs where significant structural damage to monolayers due to contact metal evaporation were reported.[18–20] The simple contact resistance model cannot capture two important phenomena that occur in contacts to 2D semiconductors: a) band bending due to the presence of a barrier (e.g., Schottky) must extend laterally into the channel,[7,21,22] and b) doping the access regions of 2D FETs is known to reduce their measured $R_c$.[9,10,23,24] So far, studies have explained the dependency of $R_c$ in these phenomena by suggesting a two-path injection mechanism,[21,25,26] but there has yet to be an implementation of these effects into the contact resistance model.

In this study we present a novel experimental analysis of contacts to monolayer $MoS_2$ which uncovers the missing component of the classical contact resistance model. We offer a correction to the model, modifying it to accurately detail contacts to atomically thin 2D semiconductors. Our thorough examination of the contacts was carried out with transfer length method (TLM), contact-end resistance (CER), and four-point probe (4PP) measurements.[16] The combination of these three techniques allowed, for the first time, the characterization and separation of both the vertical and lateral current injection resistances. Our results suggest that in addition to $\rho_c$ and $R_{sk}$, the intrinsic resistance components of the metal-semiconductor interface, a lateral resistance component must be introduced to explain the total contact resistance to 2D materials. We adopt the concept of the junction resistance ($R_{jun}$) proposed by Venica *et al.*[27] and expand it to include the lateral current injection resistance. The junction resistance accounts for the property difference between the semiconductor in the channel and under the contacts, as well as the injection over or through the Schottky barrier extending into the channel. Therefore, $R_{jun}$ encompasses the contribution of the two factors, directly connecting access region doping and contact metal variation to $R_c$. $R_{jun}$ is found to dominate contacts to atomically thin channels and must be optimized when reducing the total $R_c$ in atomically thin 2D FETs. The implementation of $R_{jun}$ into the corrected contact resistance model universally explains recent state-of-the-art data reported in the literature, unifying the physical origin of the contact resistance reduction of the various contact engineering methods.

The most common method of contact characterization is the TLM measurement, which is traditionally used to separate $R_{sh}$ from $R_c$, but does not provide information about the distribution between $\rho_c$ and $R_{sk}$ without assuming that $R_{sk} = R_{sh}$,[17] which should not be the case for contacts to 2D channels, as discussed above. To circumvent this assumption, we utilize CER measurements which in conjunction with data from TLM measurements allow us to independently characterize $R_c$, $R_{sh}$, $R_{sk}$, $\rho_c$, and $R_{jun}$. A schematic representation of a CER device structure is given in Figure 1b. We grew monolayer (1L) MoS$_2$ by chemical vapor deposition[28] (CVD) on SiO$_2$ (90 nm) on Si (p$^{++}$), and evaporated Ni as contacts. The highly doped Si substrate serves as a global back-gate, and the Methods section provides for more information. CER measurements are performed by applying current $I_{12}$ between terminals 1-2 and probing voltage $V_{23}$ between terminals 2-3, the measured contact-end resistance is then $R_{ce} = V_{23} / I_{12}$.

An important parameter of contacts is the average length which carriers travel inside the semiconductor before they are vertically injected into the metal,[16,29] named the effective transfer length ($L_{tk}$) and illustrated in Figure 1a. The brunt of the carrier injection takes place near the contact-front (i.e., the channel side of the contact), and exponentially decays deep under the contact, with a characteristic length $L_{tk}$. Therefore, in structures where the physical contact length ($L_c$) is on the same order of $L_{tk}$, there is significant current near the contact-end (i.e., the side of the contact opposite to the channel), which results in a measurable voltage drop $V_{23}$ and meaningful $R_{ce}$. Hence, $V_{23}$, and more importantly $R_{ce}$, are expected to decrease as $L_c$ is increased,[16,30] and more information is available in Supporting Section S1. The effective transfer length is given by

$$L_{tk} = \sqrt{\rho_c/R_{sk}} \tag{1}$$

and is thus a product of the intrinsic parameters of the contact, $\rho_c$ and $R_{sk}$. The contact-end resistance is given by

$$R_{ce} = \frac{\sqrt{R_{sk}\rho_c}}{\sinh(L_c/L_{tk})} = \frac{\rho_c}{L_{tk}\sinh(L_c/L_{tk})} \tag{2}$$

and can therefore be used to extract $R_{sk}$ and $\rho_c$ as previously mentioned. Figure 1c shows an atomic force microscopy (AFM) image of our CER measurement structure where the contacts are equally spaced ($d = 350$ nm) and vary in length ($L_c = 250$ nm to 700 nm). Figure 1d,e show Raman and second harmonic generation (SHG) mapping measurements that exhibit good agreement between them, confirming that most channels consist of 1L MoS$_2$, with some scattered bilayer (2L) islands. We expect the 2L areas to result in some variation in our data analysis, therefore these channels are excluded from our analysis. More details on the Raman spectroscopy are available in Supporting Section S2.

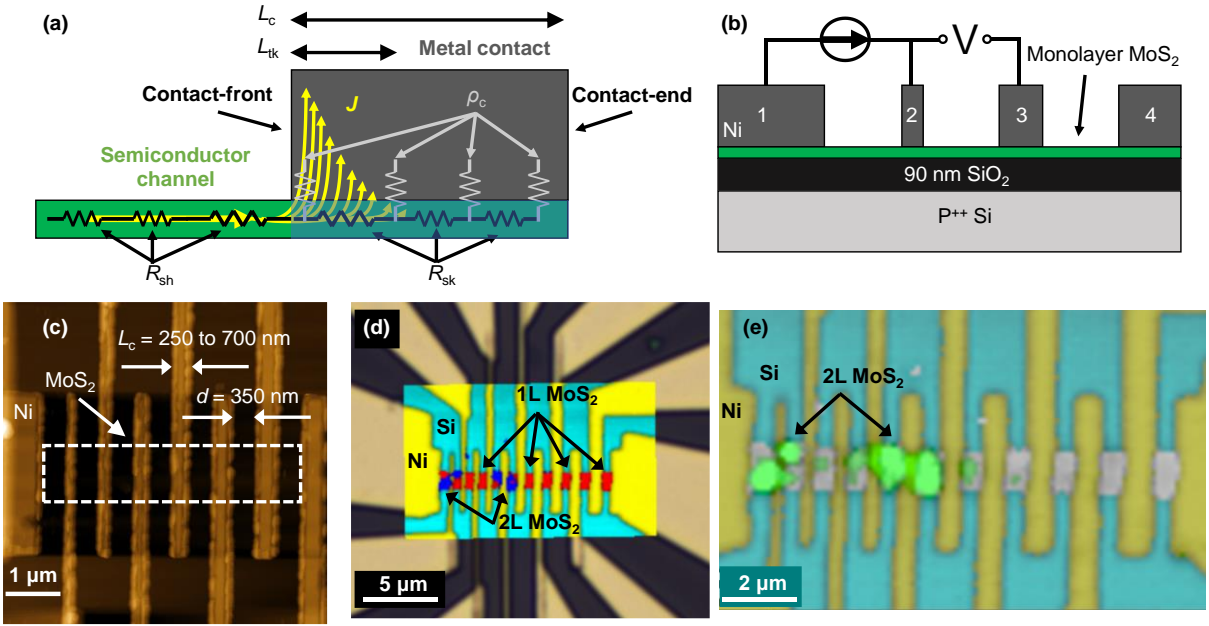

Figure 1. Transmission line model parameters and contact-end measurements structure. (a) Resistance network according to the classic contact resistance model (note that $R_{jun}$ is not present in the classic model) along the current path from a semiconductor channel to the metal contact. $R_{sh}$ is the semiconductor channel sheet resistance, $R_{sk}$ is the semiconductor sheet resistance under the contacts, and $\rho_c$ is the specific contact resistivity. The shaded area under the contact represents the difference in properties induced by the presence of the metal. The effective transfer length is noted by $L_{tk}$, and the physical contact length is $L_c$. Yellow arrows represent current flow paths. (b) Schematic cross-section of the CER test structure, and the measurement probing setup. (c) AFM image of the fabricated structures with equally distanced contacts ($d$ = 350 nm) and varying contact lengths ($L_c$ = 250 nm to 700 nm). (d) Raman and (e) second harmonic generation mapping of the same CER structure overlaid on top of an optical microscopy image. Most of the channel is 1L with some 2L islands, which should result in some variation in the data analysis.

Figure 2a shows $I_D$ vs. $V_{GS}$ measurements of a 1 µm long channel from a TLM array, performed in vacuum ambient condition (< $10^{-4}$ mbar) at room temperature and at 80 K, presented in linear and log scale. These characteristics show typical back-gate control with $I_{on} / I_{off} \approx 10^7$, and 43 µA/µm current at $V_{GS}$ = 40 V and $V_{DS}$ = 1 V at room temperature. We measured no notable hysteresis, and the gate leakage current was negligible ($I_G$ < 5 pA). To account for the variance in electrical properties between different channels in the TLM arrays, the threshold voltages ($V_{th}$) were extracted from the linear regime of the $I_D$-$V_{GS}$ plot for each channel (not shown). All the following measurements were then performed for the same back-gate overdrive voltage ($V_{GS} - V_{th}$).

$I_D$ vs. $V_{DS}$ measurements of the same channel are presented in Figure 2b, showing Ohmic characteristics at room temperature and a slight deviation at 80 K. Figure 2c,d present the total resistance ($R_{tot}$) vs channel length ($d$) for FETs from TLM arrays at room temperature and at 80 K. The TLM structure and $I$-$V$ curves used to extract the total channel resistances are included in Supporting Section S3. Note that due to low lithography yield these results include channel resistances from multiple TLM arrays from different parts of the sample, therefore the large spread in the measured data reflects the processing and property variations in different areas of the CVD-grown $MoS_2$. Extracted from the slope and $y$-intercept respectively, $R_{sh}$ and $R_c$ show dependency on the back-gate voltage $V_{GS}$, as typically observed in these structures.[8,10] The extracted sheet resistance is lower at 80 K ambient temperature due to the increase in carrier mobility, also noticeable from the linear regime slopes of the $I_D$ vs. $V_{GS}$ plots. It is noted that for 80 K the total resistance is dominated by the contact resistance $R_c$. The large spread seen in Figure 2d can thus be attributed to the variation in contact quality for the measured channels, severely hindering our ability to determine the sheet resistance from TLM.

Figure 2e shows $R_{ce}$ vs. $L_c$ where a linear behavior in log scale is observed at both room temperature and cryogenic conditions. $V_{23}$ vs. $I_{12}$ plots are shown in Supporting Section S4, and Figure 2f displays the extracted $R_{ce}$ vs. $V_{GS}$ for different contact lengths $L_c$ in the same CER array (the connecting lines between data points act as guides for the eyes). Interestingly, $R_{ce}$ (and consequently $R_{sk}$ and $\rho_c$) is found to be independent on $V_{GS}$, in striking contrast to the ~5× decrease in $R_c$ and $R_{sh}$ for the same gate bias range measured by TLM. This behavior is attributed to Fermi level pinning by the contact metal, effectively fixing the carrier concentration in 1L MoS$_2$ under the contact and hindering the charge modulation capabilities of the back-gate.[27,31,32] We therefore cannot explain how $R_c$ is experimentally found to be dependent on $V_{GS}$ using the classical model, and we conclude that a third component of the contact resistance, other than $R_{sk}$ and $\rho_c$, must be introduced to settle the contradiction. It is this unique combination of TLM and CER measurements of atomically thin 2D semiconductor FETs that uncovers the additional component of the contact resistance, which is not included in the classical transmission line model of contact resistance.

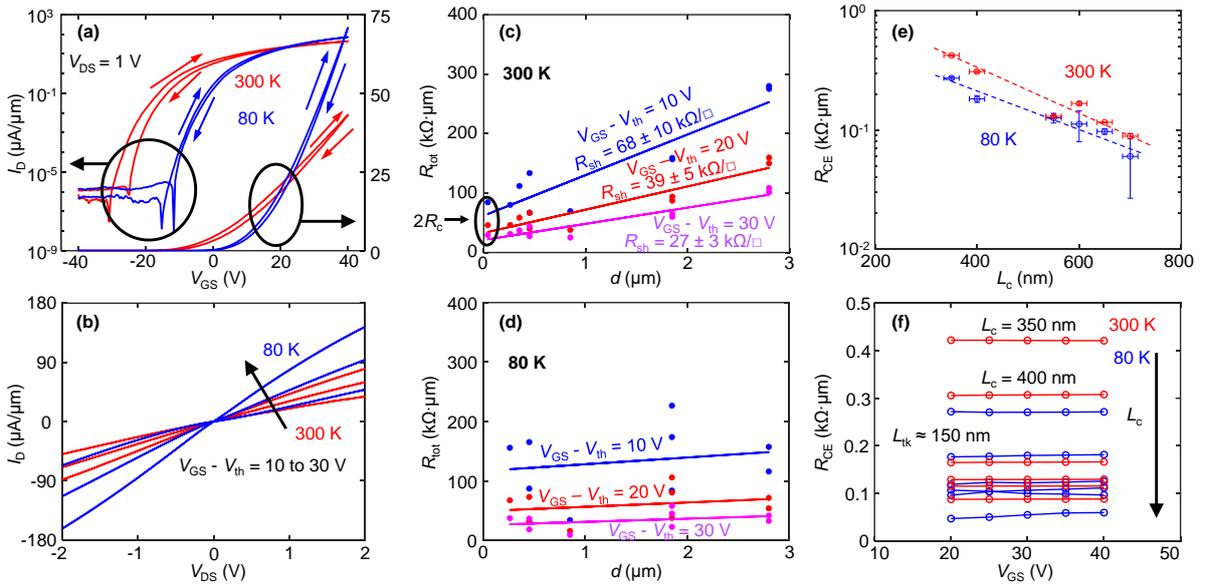

Figure 2. Electrical characteristics of MoS$_2$ FETs, TLM and CER analysis. (a) Linear and log-scale transfer and (b) output curves of a 1 μm long MoS$_2$ FET at 300 K (red) and 80 K (blue). The characteristics show typical electrical behavior with minimal hysteresis. Total resistance vs channel length at (c) 300 K and (d) 80 K. Symbols represent measured values and the solid lines are linear fits to the data. The results include data from multiple channel arrays and therefore reflect the variation in film properties due to growth and processing damage. The contact resistance extracted at 300 K is notably lower than the one extracted at 80 K and is highly dependent on the back-gate voltage. (e) Measured contact-end resistance vs contact length at 300 K and 80 K. A linear behavior in log-scale is observed, dashed lines are guides to the eye. (f) Contact-end resistance vs. back-gate voltage at 300 K and 80 K. $R_{ce}$ is independent on the back-gate voltage as was previously observed.[27]

The back-gate voltage modulates the energy bands with respect to the Fermi level in the MoS$_2$ channel, but not under the contacts as shown by the results from CER measurements. To account for the gate dependency of $R_c$ we adopt the concept of the pseudo-junction resistance ($R_{jun}$) first proposed by Venica *et al.*[27] and depicted in Figure 3. This series-resistance component of $R_c$ is attributed to the formation of a lateral junction between the channel and the area under the contact due to a variation in the material properties caused by the metal deposition. The variation can stem from several mechanisms such as physical damage from the deposition process, bandgap variation and Fermi level pinning by the contacting metal.[19,20,31–33]

For the case of Fermi level pinning near the conduction band, which is common for 1L $MoS_2$, we expect the electron concentration under the contacts to be higher than in the channel, resulting in a pseudo-junction characterized by some $R_{jun}$.[27] We further expand $R_{jun}$ to include the lateral injection resistance which describes the tunneling through, and thermionic emission over, a Schottky barrier that laterally extends from the contact to the channel, shown in Figure 3.[7,21] We note that the extreme thinness of atomically thin 2D semiconductors prevents the Schottky barrier from expanding vertically under the contact. We thus expect the majority of Schottky tunneling to take place directly into the channel, even for high concentration of charge carriers under the contact that may be induced by Fermi level pinning.[7,21] As the gate voltage increases, negative charge is induced and more electrons occupy the channel. Since the tunneling distance is inversely proportional to the carrier concentration, the lateral carrier injection becomes more probable and consequently the junction resistance decreases as the overdrive voltage increases, shown in Figure 3. We therefore conclude that $R_{jun}$ is highly dependent on $V_{GS}$. To separate between the gate-dependent and independent components of the total contact resistance ($R_{c\text{-tot}}$) we define the intrinsic resistance under the contact ($R_{c\text{-i}}$) as the combined contribution of $\rho_c$ and $R_{sk}$, thus $R_{c\text{-tot}} = R_{c\text{-i}} + R_{jun}$, as illustrated in Figure 3.

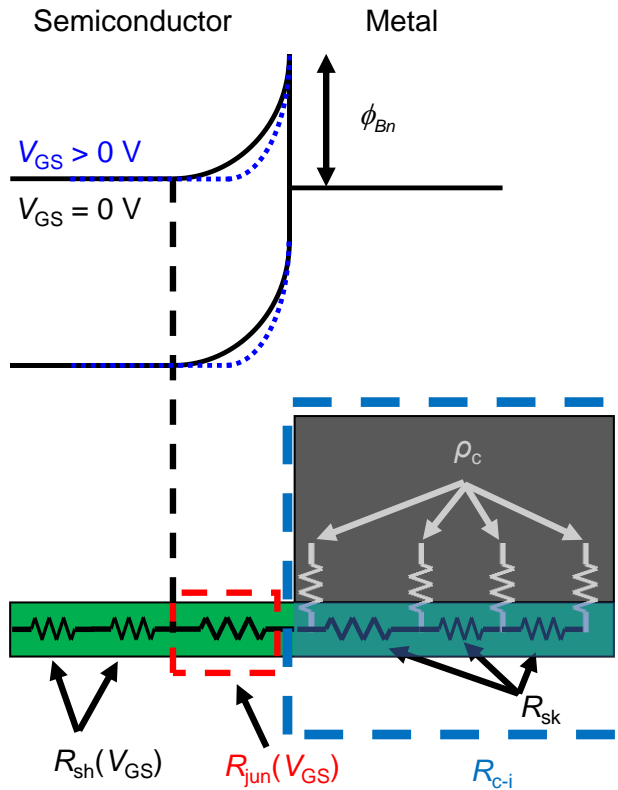

Figure 3. Cartoon illustrating the addition of the junction resistance ($R_{jun}$) to the resistance network between the metal contact and the 2D semiconductor channel. Band diagram of the Schottky barrier components of $R_{jun}$ is also presented, and the effect of the back-gate voltage is illustrated in blue. The alignment of the Schottky barrier for injection is depicted to emphasize the origin of $R_{jun}$. The intrinsic contact resistance $R_{c\text{-}i}$ is defined as the combined contribution of $R_{sk}$ and $\rho_c$.

*We characterized the different components of $R_{c\text{-}tot}$ by 4PP measurements using the device structure shown in* Figure 4Figure 4. Four point-probe (4PP) test structure and parameter extraction. (a) AFM topography image of the 4PP measurement structure. (b) *I-V* curves between terminals 1-2 of the 4PP structure for $V_{GS}$ = 0 to 40 V. The current is higher when the carrier injection (at the source electrode) takes places at the large terminal (1). $R_{c\text{-}tot}$, $R_{jun}$, and $R_{c\text{-}i}$ vs. gate overdrive voltage at (c) 300 K and (d) 80 K. Data points are presented as full round symbols and the shaded areas represent the uncertainty intervals for $R_{jun}$ and $R_{c\text{-}i}$. The square symbols show $R_{c\text{-}tot}$ obtained from TLM and the uncertainty is represented by the error bars. The TLM values are in a good agreement with the obtained 4PP results. $R_{sk}$ vs. $L_c$ using the classical contact resistance model (red) and the corrected model (blue) at (e) 300 K and at (f) 80 K.

a. In this structure the different contacts consist of two large area current driving probes (1 and 4), and two small area voltage measurement probes (2 and 3). *I-V* curves measured between large (1) and small (2) terminals, presented in Figure 4b, show reduced current when charge carriers (electrons) are injected from the latter, suggesting that the current flow bottleneck is the carrier-injecting contact. We infer that the major contribution of the contact resistance to the total channel resistance is asymmetrical and originates at the source electrode. In contacts to 2D semiconductors $R_{sk}$ and $\rho_c$ are defined by the metal-semiconductor interface, thus their resistance contribution is expected to be independent on the current injection electrode. In contrast, $R_{jun}$ includes the injection resistance that is only present where carriers are injected. Therefore, we conclude that the measured asymmetry in the *I-V* curves suggests that $R_{jun}$ is prominent at the carrier-injecting contact.

We carried out 4PP measurements by applying current $I_{14}$ between terminals 1-4 and probing the voltages $V_{12}$, $V_{23}$, and $V_{14}$ between terminals 1-2, 2-3, and 1-4 respectively. Supplementary Section S5 shows the expressions invoked to extract $R_{jun}$ and $R_{c\text{-}i}$ using the resistances measured in the 4PP setup. $R_{jun}$, $R_{c\text{-}i}$, and $R_{c\text{-}tot}$ as a function of the overdrive voltage are extracted at room temperature and at 80 K in Figure 4c,d. The values obtained from 4PP measurements (e.g., $R_{c\text{-}tot} = 25.9 \pm 4.5$ k$\Omega$ at $V_{GS} - V_{th} = 10$ V) are in good agreement with $R_{c\text{-}tot}$ from TLM ($R_{c\text{-}tot} = 22.2 \pm 11.4$ k$\Omega$ at $V_{GS} - V_{th} = 10$ V), confirming the validity of the extraction.

We found that $R_{jun}$ exhibits a strong $V_{GS}$ dependency, expectedly decreasing as the overdrive voltage, and consequently the carrier density in the channel, is increased. On the other hand, $R_{c\text{-}i}$ does not change with the gate bias or temperature, echoing the effect of Fermi level pinning on the carrier concentration under the contact, and explaining why $R_{c\text{-}tot}$ is gate-dependent while $R_{ce}$ is gate-independent. Furthermore, we see that $R_{jun}$ is the dominant contributor to $R_{c\text{-}tot}$ in our undoped devices, the lowest value is found to be $R_{jun} / R_{c\text{-}i} \approx 4.5$ for $V_{GS} - V_{th} = 35$ V. Comparing measurements at 300 K and 80 K we see that the total contact resistance is increased at lower temperature, in agreement with the TLM measurements, and that most of the variation is attributed to $R_{jun}$. This is explained by the temperature dependence of the current injection mechanism, whether thermionic emission or thermionic field emission. Therefore, reducing the temperature increases the (junction) resistance. $R_{c\text{-}i}$ remains mostly unaffected as the intrinsic properties are determined by the properties of the Ni-MoS$_2$ interface.

We note that during the CER measurements $R_{sk}$ values were extracted for different contact lengths at $V_{GS} - V_{th} = 30$ V using the classical contact resistance model at 300 K and 80 K, as seen in Figure 4e,f (red data points, $R_{jun} = 0$). An unphysical dependency on $L_c$ is observed, and the values are high compared to the channel sheet resistance $R_{sh}$ which is unexpected if Fermi level pinning is considered to induce high charge density under the contacts. However, after the implementation of the model correction with $R_{jun} = 4.5R_{c\text{-}i}$ ($R_{jun} = 6.5R_{c\text{-}i}$) at 300 K (80 K) the unexplained $L_c$ dependency vanishes for $R_{sk}$ (blue data points) as the variance the extracted values is greatly reduced. Furthermore, using the corrected model $R_{sk}$ is found to be lower than $R_{sh}$, validating the assumption of higher carrier concentration under the contacts induced by Fermi level pinning near the conduction band. We conclude that the model correction helps eliminate the unexplained behavior in the extracted values, and that overall, the combination of TLM, 4PP, and CER help determine $R_{jun}$ and show that it is the dominant component of contact resistance to undoped 1L MoS$_2$ FETs.

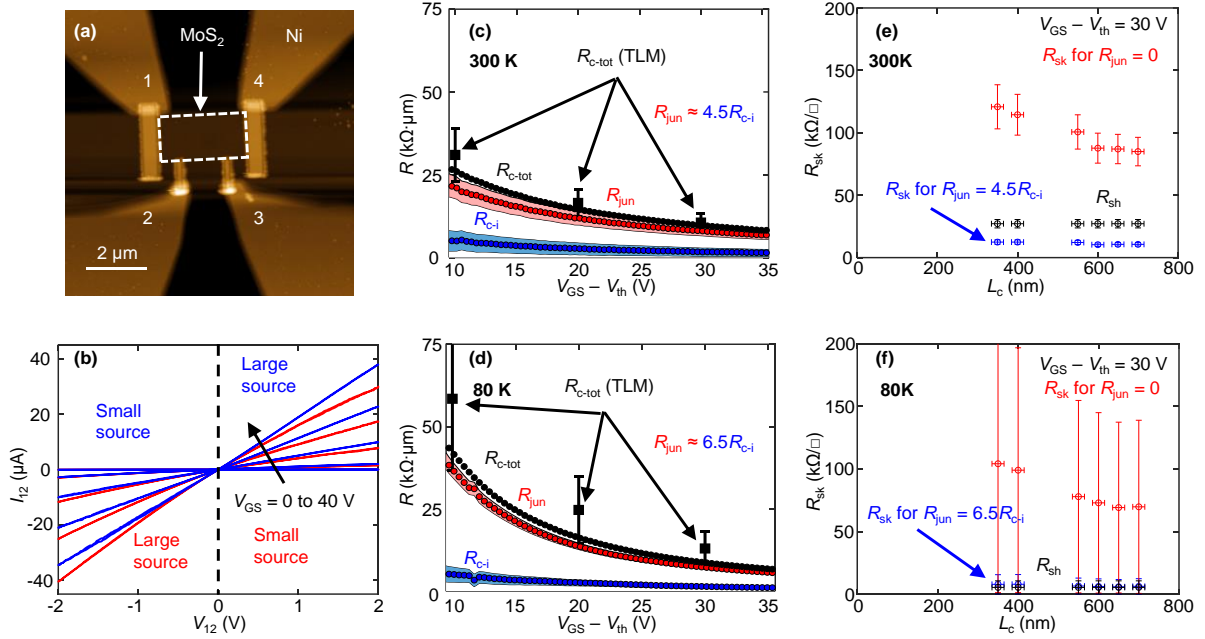

Figure 4. Four point-probe (4PP) test structure and parameter extraction. (a) AFM topography image of the 4PP measurement structure. (b) I-V curves between terminals 1-2 of the 4PP structure for $V_{GS}$ = 0 to 40 V. The current is higher when the carrier injection (at the source electrode) takes places at the large terminal (1). $R_{c\text{-tot}}$, $R_{jun}$, and $R_{c\text{-i}}$ vs. gate overdrive voltage at (c) 300 K and (d) 80 K. Data points are presented as full round symbols and the shaded areas represent the uncertainty intervals for $R_{jun}$ and $R_{c\text{-i}}$. The square symbols show $R_{c\text{-tot}}$ obtained from TLM and the uncertainty is represented by the error bars. The TLM values are in a good agreement with the obtained 4PP results. $R_{sk}$ vs. $L_c$ using the classical contact resistance model (red) and the corrected model (blue) at (e) 300 K and at (f) 80 K.

In conclusion, we have conducted TLM, CER, and 4PP measurements at room temperature and 80 K to rigorously characterize the contact resistance to 1L MoS$_2$. The TLM extraction of $R_{sh}$ and $R_c$ showed that the contact resistance is strongly dependent on the back-gate voltage. In contrast, CER results indicated that the components of the contact resistance physically located at the metal-2D semiconductor interface, $\rho_c$ and $R_{sk}$, are unaffected by gate bias which was attributed to Fermi level pinning under the contact. We proposed that in addition to the intrinsic contact resistance components $\rho_c$ and $R_{sk}$, lumped as $R_{c\text{-i}}$, a third component, $R_{jun}$, must be introduced to explain the total contact resistance to 2D semiconductors. $R_{c\text{-i}}$ accounts for the intrinsic metal-semiconductor interface resistance. $R_{jun}$ accounts for the pseudo-junction resistance and the thermionic emission over (or tunneling resistance through) the Schottky barrier at the contact edge, both extending into the channel at the carrier injecting electrode side (source). We performed 4PP measurements to distinguish between $R_{jun}$ and $R_{c-i}$, showing that $R_{jun}$ expresses the gate and temperature dependency of the total contact resistance, while $R_{c-i}$ was found to be mostly gate- and temperature-independent. Furthermore, $R_{jun}$ dominated the contact resistance to undoped channels, being at least 4× larger than $R_{c-i}$ for the tested gate bias range. $R_{jun}$ can also help explain recent reported data of state-of-the-art contact resistance reduction methods such as oxide doping[10] and semi-metallic contacts[11]. The first is explained by the increased carrier concentration near the contacts, thus lowering the tunneling distance and reducing the lateral injection resistance. The latter is explained by the lowering of the Schottky barrier for injection, again reducing the injection resistance. Therefore, by deepening the understanding of the mechanism for contact resistance reduction our findings help characterize, analyze, and ultimately design better contacts to atomically thin semiconductors.

## Methods

MoS$_2$ device fabrication

MoS$_2$ was grown by chemical vapor deposition (CVD) process on SiO$_2$ (90 nm) / Si (p$^{++}$) substrates which serve as global back-gate, more details on the CVD process are found in Supporting Section S6. All the patterning lithography steps are done by e-beam lithography using Raith EBPG 5200. First, 50 nm thick Ni contacts were e-beam evaporated under high vacuum (~4×10$^{-8}$ Torr) followed by lift-off in acetone and cleaning in IPA. Second, MoS$_2$ was etched in RIE Plasma-Therm 790 to define the channels using 90 seconds O$_2$ plasma with 20 sccm gas flow, 10 W radio frequency (RF) power, and 20 mTorr ambient pressure. Finally, Ti/Ni/Au (15/15/20 nm) thick pads and leads were evaporated with Ti acting as an adhesion layer. The TLM structures consist of equal length contacts ($L_c$ = 750 nm) with varying spacings ($d$ = 40 nm to 4800 nm). In the CER structure the contacts are equally spaced ($d$ = 350 nm) and vary in length ($L_c$ = 250 nm to 700 nm). The 4PP structure consist of two large current probes spaced 2.7 µm apart, and 2 voltage probing probes spaced 1.15 µm apart and 330 nm from the current probes.

Characterization

The MoS$_2$ sample topography was first characterized using optical microscopy (Zeiss Axiotron). The Raman and SHG spectroscopy and mapping were performed with a WITec alpha300 R instrument using 532 nm laser, 1200 g/mm grating and 100× objective lens. The WITec Project FIVE software was used for analysis, and the Si peak position at $520\ cm^{-1}$ was used for intensity normalization for all the spectra. All the electrical characterizations were carried out with a Keysight B1500 semiconductor parameter analyzer (SPA) in a JANIS probe station in vacuum conditions ($<10^{-4}$ mbar) at room temperature (near 300 K) and at 80 K.

## Acknowledgments

The fabrication was performed at the Technion Micro-Nano Fabrication & Printing Unit (MNF&PU) with support from the Russell Berrie Nanotechnology Institute (RBNI). We thank Guy Ankonina for his help with the SHG measurements. This work was supported in part by ISF grant # 1179/20. R.W.G. acknowledges partial support from the National Science Foundation Graduate Research Fellowship Program under Grant DGE-1656518 and from Intel. E.P. acknowledges partial support by ASCENT, one of six centers in JUMP, a SRC program sponsored by DARPA.

## Additional information

The Supporting Information is available free of charge at

Contact-end modeling information; Raman, PL, and SHG data of monolayer $MoS_2$ devices used for mapping; Device structure and *I-V* curves used for the TLM analysis; *I-V* curves used for the CER measurement; Equations used for the extraction of $R_{c\text{-}i}$ and $R_{jun}$ from 4PP measurements; CVD growth of monolayer $MoS_2$.

## AUTHOR INFORMATION

### Corresponding Author

*Eilam Yalon – Viterbi Faculty of Electrical & Computer Engineering, Technion – Israel Institute of Technology, Haifa 32000, Israel

Email: eilamy@technion.ac.il

### Authors

Emanuel Ber – Viterbi Faculty of Electrical & Computer Engineering, Technion – Israel Institute of Technology, Haifa 32000, Israel

Ryan W. Grady – Electrical Engineering, Stanford University, Stanford, California 94305,

United states

Eric Pop – Electrical Engineering and Materials Science and Engineering, Stanford University, Stanford, California 94305, United States

**Author Contributions**

E.B. and E.Y. conceived the experiments and wrote the manuscript with input from all authors. R.W.G. grew the MoS$_2$. E.B. fabricated the devices and performed electrical, PL, and Raman characterizations.

**Notes**

The authors declare no competing financial interest.

Emanuel Ber[1], Ryan W. Grady[2], Eric Pop[2,3], and Eilam Yalon[1,*]

[1]*Viterbi Faculty of Electrical & Computer Engineering, Technion-Israel Institute of Technology, Haifa-32000, Israel*
[2]*Department of Electrical Engineering, Stanford University, Stanford, CA 94305, USA*
[3]*Department of Materials Science & Engineering, Stanford University, Stanford, CA 94305, USA*
[*]*E-mail:* eilamy@technion.ac.il

# Table of Contents

## S1. Contact-end modeling information

We use contact-end resistance (CER)[1] measurements to obtain information on the intrinsic properties of current injection in the metal-semiconductor interface. A depiction of the test structure is shown in Figure S5a, where contacts of increasing lengths ($L_c$) are equally distanced from one another. It is important to note that the contact length is designed to be shorter or near the effective transfer length ($L_{tk}$). The current flow under a short contact ($L_c < L_{tk}$) is presented in Figure S5b, where it is shown to exponentially decay along the contact. This decrease is dictated by the relation between the two intrinsic properties of the metal-semiconductor interface, the sheet resistance under the metal contact ($R_{sk}$) and the specific contact resistivity ($\rho_c$). Therefore, by characterizing the exponential decay of the current under the contact the CER measurement provides crucial information regarding the contact interface.

To perform CER measurements current is applied between two adjacent contacts (1 and 2), while a voltage drop is measured at a third auxiliary contact (3). Because contact 2 in the CER structure is designed to be short, the current does not completely vanish near its contact-end resulting in a measurable potential difference between contacts 2 and 3. In other words auxiliary contact 3 is used to probe the potential at the contact end of terminal 2. This voltage drop between contacts 2 and 3 is proportional to the amount of leftover current at the contact-end, which decreases as $L_c$ is increased relative to $L_{tk}$. Therefore, to measure meaningful voltage at the auxiliary terminal two approaches are taken: a) driving high currents and b) fabricating the shortest contacts possible within the lithography and processing limitations. Dividing the measured voltage by the applied current yields the contact-end resistance ($R_{ce}$), and plotting it vs. $L_c$ is expected to result in a linear graph in log-scale as depicted in Figure S5c.[2]

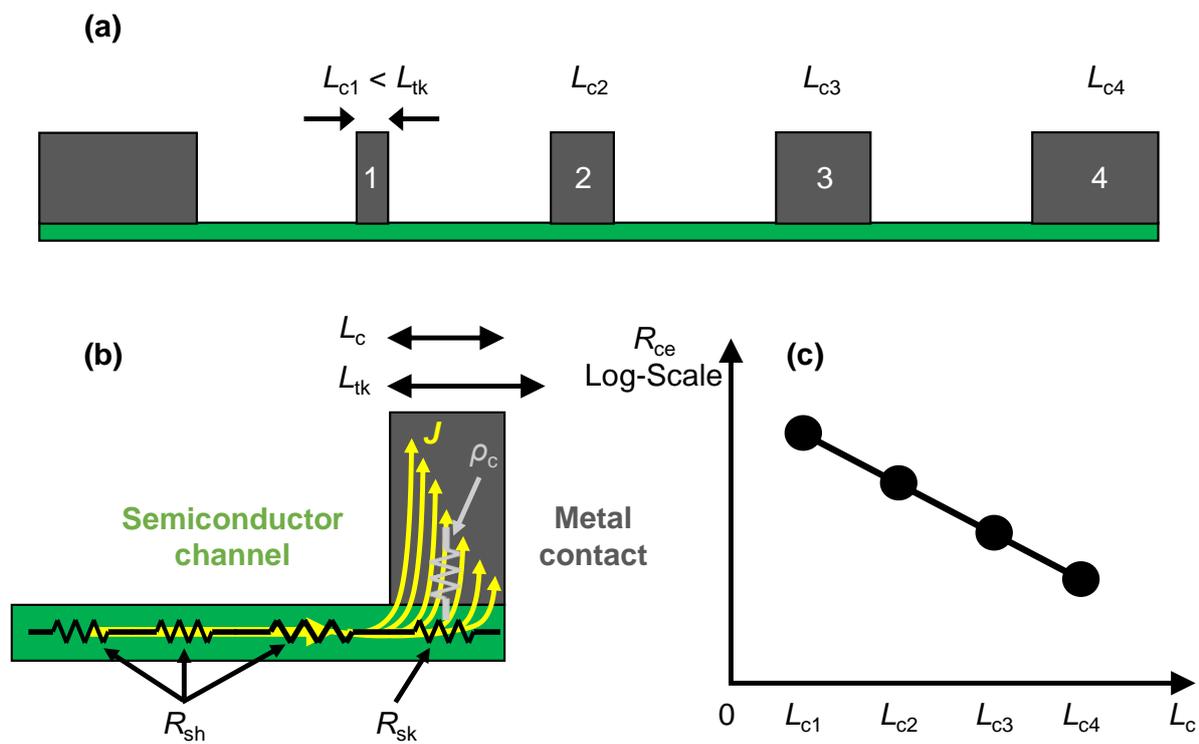

Figure S5. Schematics of contact-end resistance measurement structure, current distribution under the contact and typical measurement results. (a) Cartoon showing multiple contacts with varying lengths used as part of the contact-end measurements. To improve the accuracy of the measurement, contacts acting as terminal 2 should be comparable to the transfer length $L_{tk}$. (b) Depiction of the current flow under a contact shorter than $L_{tk}$. In this case there is significant current at the contact-end. (c) Expected results of $R_{ce}$ vs. $L_c$ plot. A linear behavior in log-scale corresponds to the exponential decay of the current under the contact.

## S2. Raman, PL, and SHG data of MoS₂ devices used for mapping

Raman spectroscopy, photoluminescence (PL), and second harmonic generation (SHG) measurements were used to map the channel material (MoS$_2$) and number of layers in our test structures. Individual Raman and PL spectra of monolayer (1L) and bilayer (2L) MoS$_2$ are shown in Figure S6a,b. The Raman spectra were normalized by the Si peak intensity (from the substrate), and the PL spectra were normalized by the maximum intensity of the 1L MoS$_2$ measurement. We then performed an area scan of the device structure and used the spectra shown in Figure S6a for the True Component Analysis offered by WITec Project FIVE software to assign a layer number to each point on the map.

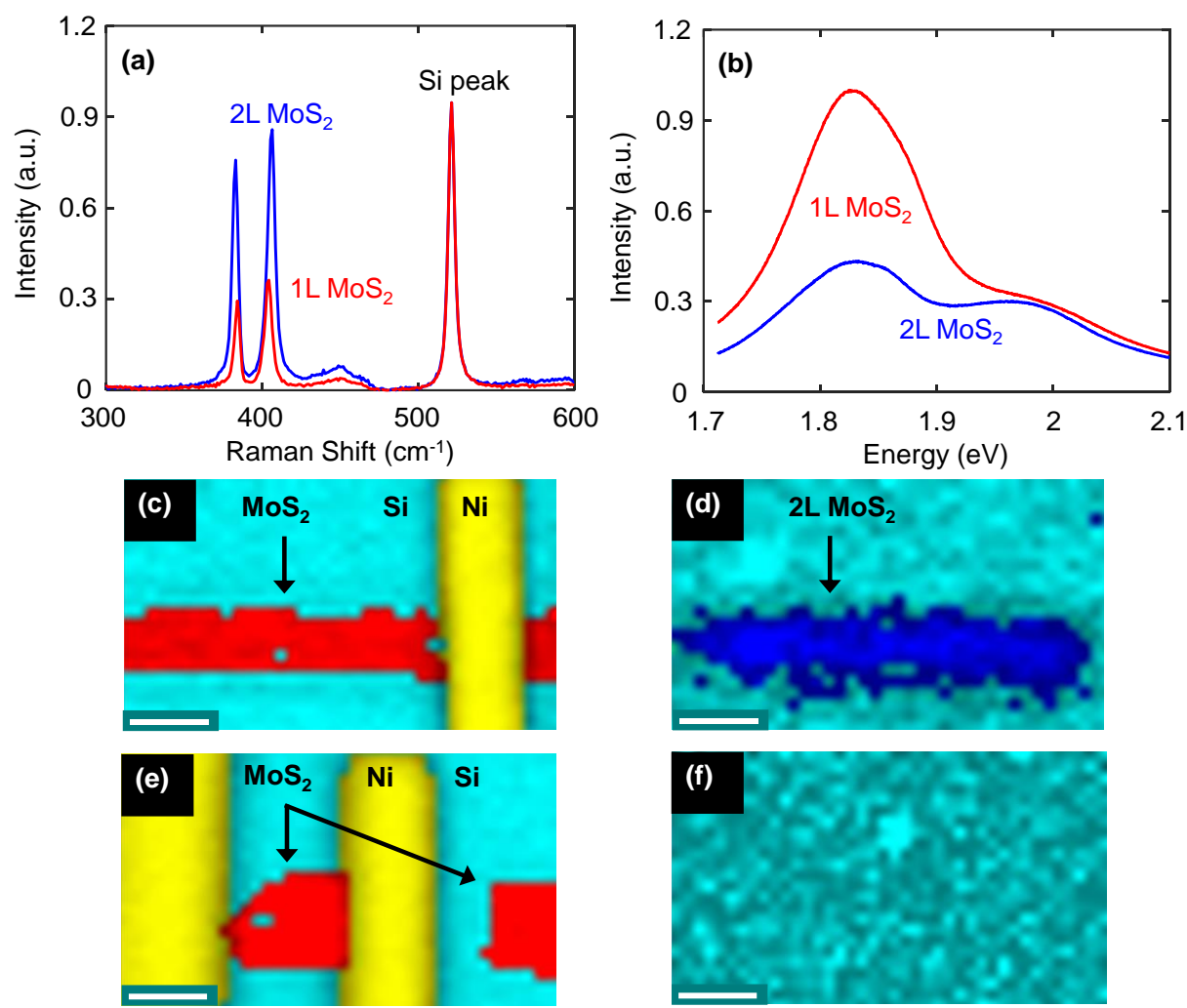

Figure S6. Raman, PL, and SHG spectra and mapping of MoS$_2$. (a) Typical Raman spectra of 1L and 2L MoS$_2$. The measured data is normalized by the intensity of the Si peak. (b) PL spectra of 1L and 2L MoS$_2$. The 2L MoS$_2$ spectrum is suppressed in comparison to the spectrum of 1L MoS$_2$. (c) Raman and (d) SHG mapping of 2L MoS$_2$ channel. The Raman mapping is insufficient in determining whether the MoS$_2$ is 1L or 2L, and the SHG mapping reveals the 2L nature of the material. (e) Raman and (f) SHG mapping of 1L MoS$_2$ channel. The Raman mapping showing the presence of MoS$_2$ but there is no signal from SHG which determines the 1L nature of the material. All spectra and mapping were taken at room temperature with a 532 nm laser for Raman and PL and 1064 laser for SHG. The scale bar for all figures is 500 nm long.

We performed SHG mapping to further differentiate between 1L and 2L areas as shown in Figure S6c-f. In Figure S6c,d we can see that the Raman mapping alone is insufficient in determining whether the channel consists of 1L or 2L MoS$_2$, but the SHG mapping shows that it is 2L. In Figure S6e,f however, there is no SHG signal from the mapped area therefore we conclude that the MoS$_2$ in these channels is 1L.

## S3. Device structure and *I-V* curves used for the TLM analysis

Transfer length method (TLM) measurements are performed by measuring the total resistance for semiconductor channels of increasing lengths, as shown in Figure S7a. The channel lengths in our TLM structures varied from 40 to 4800 nm, while $L_c$ remained constant at 500 nm. An *I-V* curve of a single 1.85 μm-long channel for various back-gate voltages is presented in Figure S7b, showing a linear behavior. The slope of the curve is used to extract the total channel resistance, which shows a dependency on the applied back gate bias.

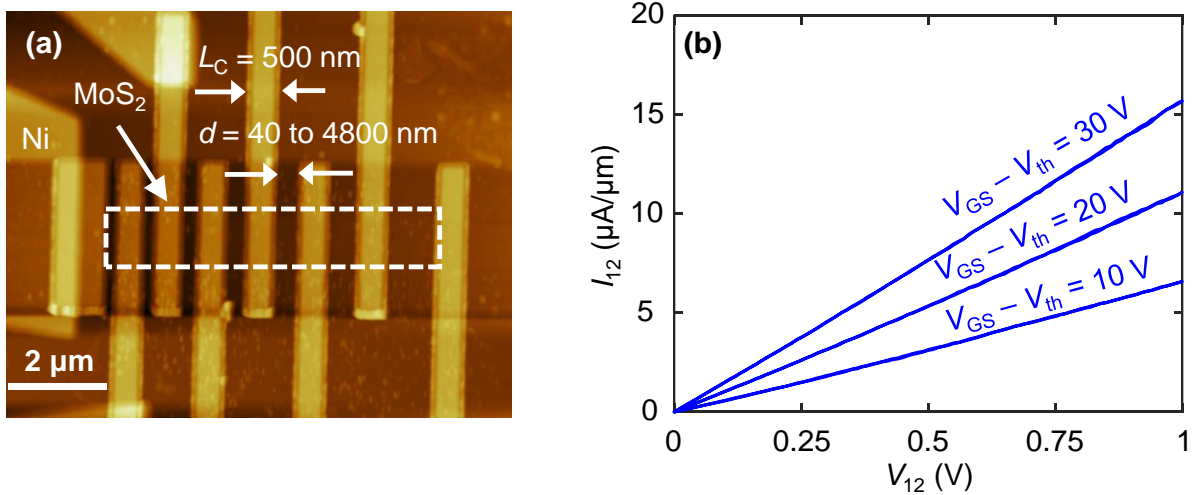

Figure S7. TLM measurement structure and typical measurement result. (a) Atomic force microscopy image of the TLM structure used in our analysis. The MoS$_2$ layer is pointed out by the dashed white outline. (b) *I-V* curve of a single 1.85 μm-long channel at various back gate voltages. $V_{th}$ is the threshold voltage of each channel.

## S4. *I-V* curves used for the CER measurement

CER measurements are performed by extracting $R_{ce}$ for contacts of increasing lengths using structures depicted in Figure 1c from the main text. The measured voltage at an auxiliary contact ($V_{23}$) vs the current applied between two adjacent contacts ($I_{12}$) for various back-gate voltages is presented in Figure S8a,b. The length of the current injection contact (2) is 250 nm, the driving contacts, 1 and 2, are spaced 350 nm apart and the highest applied voltage is $V_{12} = 1$ V. As reported in the main text, $R_{ce}$ (the slope of the curve) is independent on the applied gate bias. To measure significant voltage at the auxiliary contact moderately high current must be applied, therefore the back-gate bias range is limited by the channel resistance at the lower end (too few carriers) and by the oxide leakage at the upper end.

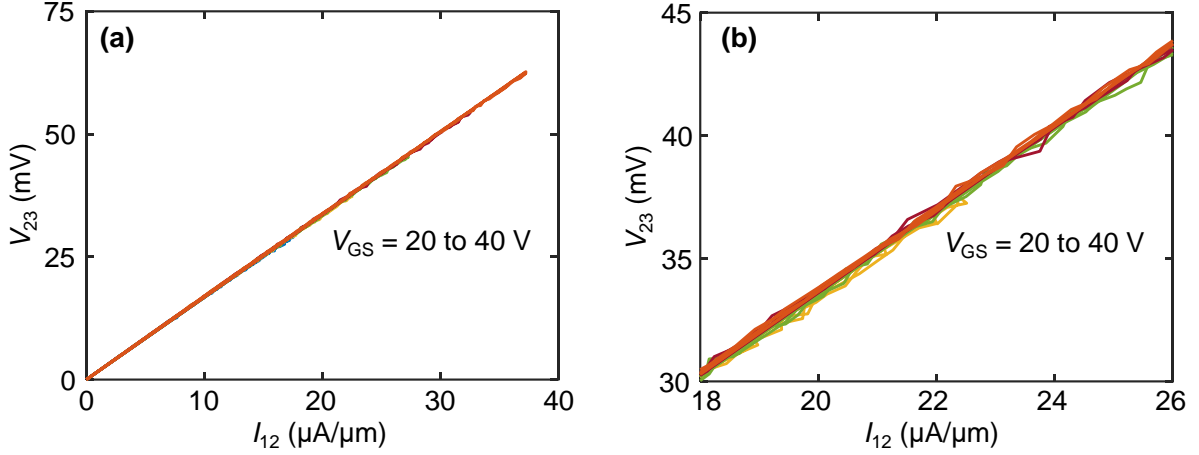

Figure S8. *I-V* curves used for CER measurements in (a) full and (b) limited range. $R_{ce}$ is extracted from the slope of the line.

## S5. Equations used for the extraction of $R_{c-i}$ and $R_{jun}$ from 4PP measurements

To extract the values of the intrinsic contact resistance ($R_{c-i}$) and the junction resistance ($R_{jun}$) we performed four-point probe (4PP) measurements as described in the main text and used the following formulation to extract the individual resistances. First, the resistance between the outer current driving probes is expressed by

$$WR_{14} = 2R_{c-i} + R_{jun} + R_{sh}d_{14} \qquad (S1)$$

Where $R_{14}$ is the resistance between probes 1 and 4, $R_{sh}$ is the sheet resistance of the channel separating the probes, and $d_{14}$ is the distance between the two probes. Because the current path runs through two semiconductor-metal interfaces $R_{c-i}$ is taken into consideration twice. As seen in the main text, $R_{jun}$ is only prominent at the current injecting probe and thus appears only once in the above expression. Next, the resistance between the current injecting electrode and the voltage probe is expressed as

$$WR_{12} = R_{c-i} + R_{jun} + R_{sh}d_{12} \quad (S2)$$

Where $R_{12}$ is the resistance between probes 1 and 2, and $d_{12}$ is the distance between the two probes. Since probe 2 is defined as a voltage probe (high Z) it does not drive any current and therefore $R_{c-i}$ is present only once in the above expression. Finally, the resistance between the voltage probes is expressed as

$$WR_{23} = R_{sh}d_{23} \quad (S3)$$

Where $R_{23}$ is the resistance between probes 2 and 3, and $d_{23}$ is the distance between the probes. As both probes do not drive any current, only the sheet resistance between them is considered. From equation S3 we can extract the sheet resistance of the channel. The distance $d_{23}$ is measured by atomic force microscopy and $R_{23}$ is extracted by dividing the voltage measured between probes 2 and 3 ($V_{23}$) by the total current running through the channel ($I_{12}$). $R_{c-i}$ and $R_{jun}$ are then extracted by using equations S1, S2, and S4, resulting in the expressions

$$R_{c-i} = W(R_{14} - R_{12}) - WR_{23}\left(\frac{d_{14} - d_{12}}{d_{23}}\right) \quad (S5)$$

$$R_{jun} = WR_{12} - R_{c-i} - WR_{23}\frac{d_{12}}{d_{23}}. \quad (S6)$$

Expressions S5 and S6 show explicitly how the two resistances can be individually extracted, as discussed in the main text, and shown in Figure S4c,d of the main text.

## S6. CVD growth of monolayer MoS$_2$

Monolayer MoS$_2$ was grown by chemical vapor deposition using solid source precursors. First, a p$^{++}$ Si substrate with 90 nm of oxide was prepared by coating with hexamethyldisilazane (HMDS) to render the oxide hydrophobic. Then, approximately 30 μL of 100 μM aqueous solution of perylene-3,4,9,10-tetracarboxylic acid tetrapotassium salt (PTAS), a seeding promoter, was deposited around the edges of a substrate and subsequently evaporated on a hot plate. A small quantity, under ~500 μg, of MoO$_3$ (Alfa Aesar, 99.9995% purity), was placed in a crucible with the substrate face down over the crucible. This was placed in a 2" tube furnace downstream from a precursor boat containing sulfur (Alfa Aesar, 99.9995% purity). The growth was carried out under argon ambient at 30 sccm. The furnace was heated to 850 °C at a ramp rate of 27.5 C min$^{-1}$, held at 850 °C for 15 minutes, and then allowed to cool naturally.